# Thick MgB$_2$ film with (101) oriented micro-crystals

Chinping Chen [1], Xinfeng Wang[1], Ying Lu[1], Zhang Jia[1], Jing-pu Guo[1],

Xiao-nan Wang[1], Meng Zhu[1], Xiangyu Xu[1], Jun Xu[2], Qing-rong Feng[1*]

1. Department of Physics and State Key Laboratory for Mesoscopic Physics,

Peking University, Beijing 100871, P.R. China

2. Electron Microscopy Laboratory, School of Physics, Peking University,

Beijing 100871,P. R. China

**Abstract**

Very thick, ~ 40 μm, clean, and highly textured MgB$_2$ film was effectively grown on an Al$_2$O$_3$ substrate. The fabrication technique is by the hybrid physical-chemical vapor deposition (HPCVD) using B$_2$H$_6$ gas and Mg ingot as the sources. The X-ray diffraction (XRD) analysis shows a highly (101)-oriented MgB$_2$ crystal structure without any impurity detected. There is no signal from the substrate in the XRD spectrum, indicating that the film thickness exceeds the X-ray penetration length. Scanning electron microscopy (SEM) reveals that the film is composed of highly-packed MgB$_2$ micro-crystals with a uniform size distribution of about 2 μm in diameter and 0.2 μm in thickness. According to the compositional analysis of energy-dispersive X-ray spectroscopy (EDX), no oxygen, hence no MgO, exists in the textured film, consistent with the XRD result. Also, the transport properties are similar to those of a single crystal, indicating a clean film of good crystallite. The zero field transition temperatures are determined as T$_C$(onset) = 39.2 K and T$_C$(zero) = 38.4 K, giving a sharp transition typical of a clean sample. The residual resistivity ratio (RRR) is determined as 6.4 and the magnetoreisitance (MR) is about 28 % at 40 K under the applied field of 9 T, which are similar to those of a single crystal. The zero temperature upper critical field, H$_{C2}$(0), is extrapolated as 19 T from the T$_C$(onset) at applied field up to 9 T.

PACS: 74.25.Ha; 74.62.Bf; 74.72.Jt; 74.76.-w

Keywords: MgB$_2$ film, upper critical field

The MgB$_2$ superconductor is a promising candidate to replace the Nb-based superconductor for the conventional low T$_C$ superconducting applications with the transition temperature of 40 K[1]. The high operating temperature, 20 ~ 30 K, makes the compact compressor-based cryo-cooler an effective and easy-operating system to provide the appropriate temperature environment without liquid helium. Its simple binary metallic structure with the cheap raw materials available makes the commercialization highly competitive. Many researches have been focused on the fundamental studies of its physical properties[2-3], while others on the application aspects[4]. Application-wise, the MgB$_2$ films are especially important not only in the superconducting circuits and devices, but also as a possible pre-stage for the preparation of the MgB$_2$ tapes and wires in the industry at lower lost and simpler process. An effective approach to produce controlled quality, thick MgB$_2$ film is therefore especially important for the purpose of electronic devices and large-scaled superconducting industrial applications.

We have grown very thick MgB$_2$ film with uniform thickness of about 40 μm on the Al$_2$O$_3$ substrate. It is based on the techniques of HPCVD using the B$_2$H$_6$ precursor and Mg ingot as the active sources, similar to that reported previously[5-6]. Additional gas mixture of H$_2$ ( 4 % ) and Ar ( 96 % ) were also flowing in the chamber, serving to reduce the oxygen content and to suppress any possible further oxidation of the sample during the thermal process. Also, the H$_2$ would cut down the decomposition speed of B$_2$H$_6$ for an easily control of the reaction rate. The flow rate of B$_2$H$_6$ was about 30 sccm at the pressure of 2 KPa and the background mixture, H$_2$ + Ar, about 300 sccm at 18 KPa. The temperature in the chamber was controlled within the range of 680 to 830 °C. Under the aforementioned conditions, the film deposition rate is calculated as 17 nm/s, which is much more effective than the result reported by Pogrebnyakov et al[7] considering the lower flow rate of H$_2$B$_6$ at a lower pressure in the present work.

The XRD was performed using a Philip x' pert diffractometer. The spectrum in Fig. 1 shows a highly (101)-oriented MgB$_2$ crystal structure without any other impurity. The (101) peak in the spectrum is so sharp that the double-peaked structure, resulting from the Cu K$_{\alpha 1}$ and Cu K$_{\alpha 2}$, is clearly resolved with a separation of 2θ = 0.1°, see the close-up in the upper-left inset of Fig. 1. Note that the relative ratio of the peak heights is exactly 2.0, consistent with the relative radiation intensity ratio of 2.0. In addition, according to the Bragg equation, $\lambda = 2d \sin\theta$, the difference in wavelength of the two Cu Kα lines would lead to the angular difference of $2\theta = 0.11^o$, in the calculation, using the lattice spacing, $d = 0.2166$ nm, corresponding to the (101) peak. The angular separation would be too small to be the (200) peak of MgO, which is usually observed in the $MgB_2$ samples. The (002) peak of $MgB_2$, which is barely visible in the major spectrum, is amplified in the upper-right inset in an expanded vertical scale. The characteristic peaks for the substrate do not appear on the spectrum because the film thickness, ~ 40 μm, exceeds the typical X-ray penetration length, a few micrometers.

The SEM observations were carried out by the LEO 1450VP and the FEI STRATA DB235 electron microscopes. The energy-dispersive X-ray (EDX) spectroscopy was performed using the INCA ENERGY300 of Oxford installed on the LEO 1450VP. Fig. 2(a) shows the cross sectional image of the grown film, as marked by the two arrows, taken at a slightly tilted angle. The film is deposited with a uniform thickness of about 40 μm over the $Al_2O_3$ substrate shown at the lower part of Fig. 2(a) in a darker shade. It is much thicker than any other $MgB_2$ film grown by the same HPCVD method reported previously[8]. The layer structure is clearly visible on the cross sectional view, indicating that it is textured. There are humps with dimensions on the order of a few tens of a μm distributed over the film surface, causing the roughness as viewed from the tilted angle. Note that there are a few the speckles or stains in a lighter shade locating over the cross-sectional view of the film and the substrate. They are possibly attributed to the contamination from the sample cutting for the SEM investigations. A typical porous hump residing over the surface of the good textured $MgB_2$ film is shown in Fig. 2(b). It has a sponge-like, porous structure. By the EDX compositional analysis, there are O, Mg, and B atoms existing in these humps. The general feature of the film surface other than the hump area is shown in Fig. 2 (c). It is mostly composed of highly-packed, micro-flake like $MgB_2$ crystals with uniform size distribution. The average size is 2 ~ 3 μm and the thickness is about 0.2 ~ 0.3 μm. By the EDX analysis, there are only Mg and B atoms in this area with relative atomic ratio close to 1:2. No trace of oxygen atom at all exists. This indicates that the micro-crystals have a high-purity $MgB_2$ phase. Direct evidences of the MgO were also observed in the shape of needle-like nanorod underneath the outer-most surface of the porous hump, see in Fig. 2 (c). The amount of the MgO is so little that it is beyond the XRD detection resolution. Hence, the analysis by the SEM and EDX is consistent with that of XRD.

The phase purity is an important issue on the superconducting properties. According to the

XRD, no impurity phases of MgO and Mg exist in the film. Neither has any evidence of the other B-rich phases such as $MgB_4$ or $MgB_7$[9], which requires a higher formation temperature. Nonetheless, it is difficult to avoid completely the MgO after long exposure to the residual oxygen in the flowing Ar at high temperature [10]. What is interesting is that the traces of the minute MgO observed by the SEM seem to exist in the porous humps over the textured $MgB_2$ film. Perhaps, this is due to the high solubility and mobility of oxygen in the $MgB_2$ phase and the oxygen migrates out to the surface easily to react with the magnesium[11].

The temperature-dependent resistivity (ρ-t), with or without the applied field, was performed by the standard 4-probe measurement using the Quantum Design PPMS system. The electrical contact was made using the silver paste. The result from 300 down to 4 K, is plotted in Fig. 3. The transition temperatures are determined as $T_C(onset) = 39.2$ K and $T_C(zero) = 38.4$ K, see the inset of Fig. 3. The corresponding transition width is as narrow as 0.8 K. This is comparable or sharper than most of the thin film or even bulk samples reported[12,13]. According to the resistivity at 300 K and 40 K, RRR is calculated as 6.4, falling in the range for single crystals [14,15]. Furthermore, the MR effect calculated from the field measurements at T = 40 K and H = 9 T, see the inset of Fig. 4, is roughly 28 %. This is similar to the value of a single crystal, ~ 25 %, at 40 K under $H_{//C} = 9$ T[16]. These transport properties are consistent with the results by the XRD analysis and the SEM observation that the fabricated thick $MgB_2$ film is very clean with high crystallinity. The upper critical field at T = 0 K, $H_{C2}(0)$, is extrapolated as 19 T, see Fig. 4, from the temperature dependent resistivity measurement under applied field up to 9 T, shown in the inset of Fig. 4. Also plotted in the same figure by the two solid curves for comparison are the $H_{C2}//c$ and $H_{C2}//ab$ for a single crystal, taken from Fig 4 of reference 16. The $H_{C2}(T)$ is close to the value of $H_{C2}//ab$ of a single crystal.

In conclusion, very thick with uniform thickness of about 40 μm, well textured, and clean $MgB_2$ film was effectively grown by the method of HPCVD. The film is mainly composed of tightly-packed, highly (101)-oriented micro-crystals, according to the XRD analysis and the SEM observations. The transport properties in the normal state, such as RRR ~ 6.4, MR ~ 28 % at 40 K under 9 T of applied field, are close to those of a single crystal reported previously. The superconducting transition occurs at $T_C(onset) = 39.2$ K, $T_C(zero) = 38.4$ K, corresponding to a sharp transition width of 0.8 K, indicating the phase purity of the film.

Acknowledgement

This research was a part project of the Department of Physics of Peking University, and is supported by the Center for Research and Development of Superconductivity in China under contract No. BKBRSF-G19990646-02.

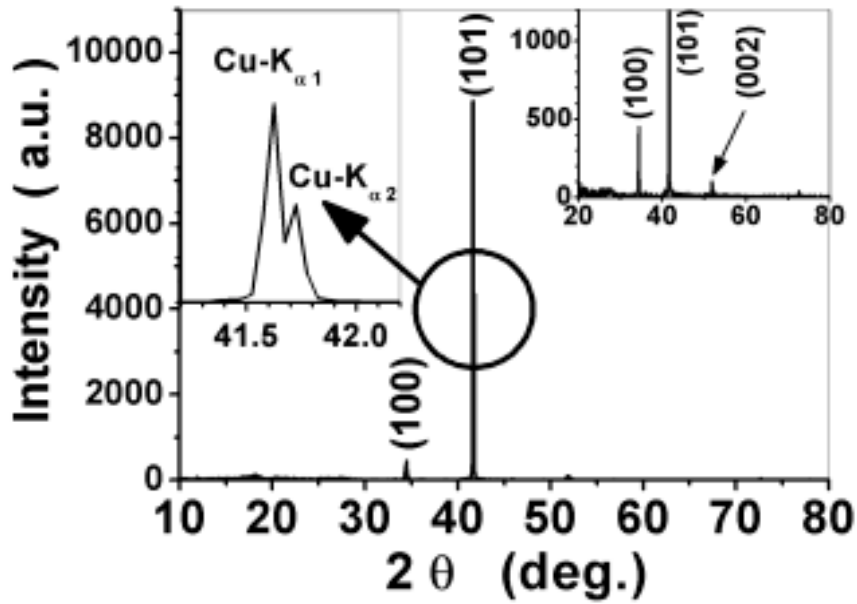

Fig. 1  XRD spectrum for the thick $MgB_2$ film. The crystal structure is highly (101)-oriented. The film thickness, ~ 40 μm, exceeds the X-ray penetration length. The (101) peak has a double-peaked structure with a separation of $2\theta = 0.1°$ and relative intensity of 2 to 1, see the inset in the upper-left, resulting from the Cu $K_{\alpha1}$ and $K_{\alpha2}$. In the upper-right inset with the Y-scale expanded by a factor 10, the (002) peak shows up, which is barely visible in the major figure. No other impurity phase has been detected in the spectrum.

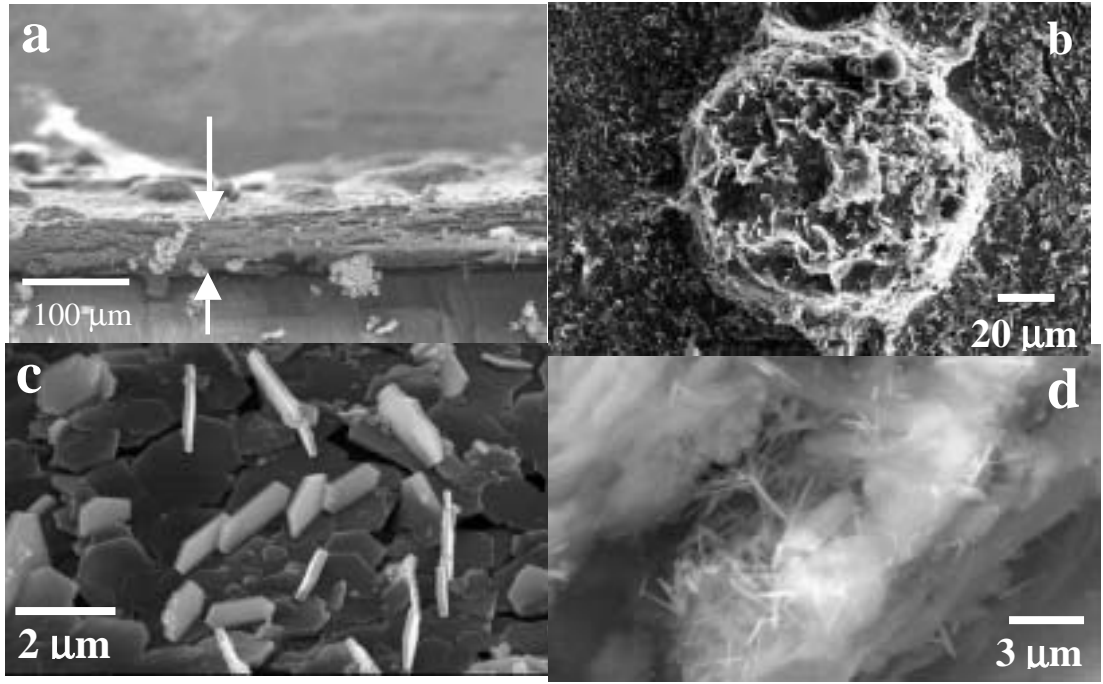

Fig. 2 SEM characterizations of the fabricated thick MgB$_2$ film. (a) Cross sectional image of the grown film taken at a slightly tilted angle. The film has a uniform thickness of about 40 μm, marked by the two arrows. The layer structure is visible on the cross sectional view, indicating that the film is textured. The Al$_2$O$_3$ substrate is at the bottom part with a darker shade. Humps are distributed over the film surface to show bumpiness. Speckles or stains appear in the figure possibly attributed to the contamination during the sample cutting for the SEM. (b) A typical porous hump residing over the surface of the well-textured MgB$_2$ film. There are oxygen exists in this structure along with magnesium and boron, according to the EDX analysis. (c) Micro-flakes of the MgB$_2$ crystals in a selective area around the hump shown in Fig. 2(b). Magnesium and boron elements exist in this area without oxygen by the EDX analysis, indicating that there is no MgO. (d) Images of MgO nanorods (nanotube) in the needle shape located underneath the outer surface of the hump shown in Fig. 2(b).

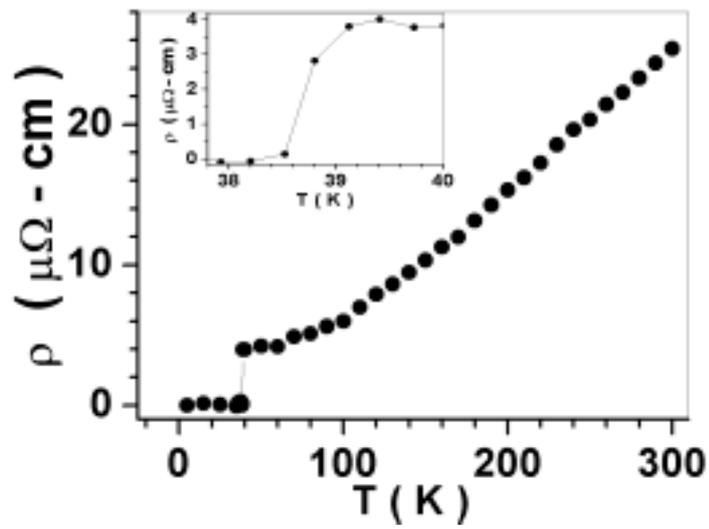

Fig. 3 Temperature-dependent resistivity of the thick MgB$_2$ film. T$_C$(onset) is 39.2 K and T$_C$(zero) is about 38.4 K. The normal state resistivity is, ρ(300K) = 25.4 μΩ-cm, ρ(40K) = 3.97 μΩ-cm. This gives RRR = 6.4.

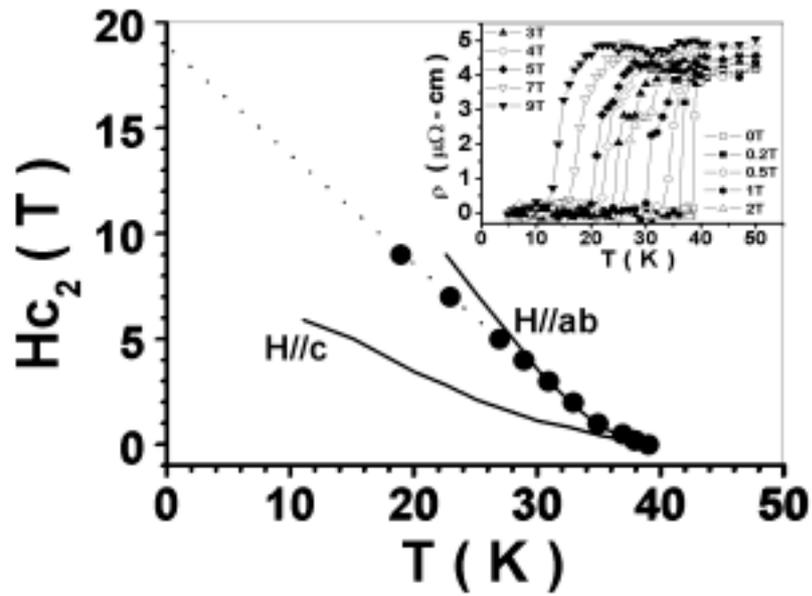

Fig. 4 $H_{C2}(T)$ determined from the ρ-T measurement under applied field up to 9 T, see the inset.. The dotted line extrapolated to T = 0 K gives $H_{C2}(0)$ ~ 19 T. The two solid curves are taken from Fig. 4 of reference [16] for the two upper critical fields, along *c*-axis and in *ab*-plane of a $MgB_2$ single crystal.